\documentstyle[aaspp4,10pt,flushrt,epsfig]{article}

\lefthead{Bildsten}
\righthead{Gravitational Radiation from Accreting Neutron Stars} 

\begin{document}

\title{Gravitational Radiation and Rotation of Accreting Neutron Stars} 

\author{Lars Bildsten} \affil{Department of Physics and Department of
Astronomy \\ 366 LeConte Hall, University of California, Berkeley, CA
94720 \\ email: bildsten@fire.berkeley.edu}

\begin{abstract}

Recent discoveries by the {\it Rossi X-Ray Timing Explorer} indicate
that most of the rapidly accreting ($\dot M \gtrsim 10^{-11} M_\odot \
{\rm yr}^{-1}$) weakly magnetic ($B\ll 10^{11} \ {\rm G}$) neutron
stars in the Galaxy are rotating at spin frequencies $\nu_s \gtrsim
250 \ {\rm Hz}$. Remarkably, they all rotate in a narrow range of
frequencies (no more than a factor of two, with many within 20\% of
300 Hz). I suggest that these stars rotate fast enough so that, on
average, the angular momentum added by accretion is lost to
gravitational radiation. The strong $\nu_s$ dependence of the angular
momentum loss rate from gravitational radiation then provides a
natural reason for similar spin frequencies. Provided that the
interior temperature has a large scale asymmetry misaligned from the
spin axis, then the temperature sensitive electron captures in the
deep crust can provide the quadrupole needed ($\sim 10^{-7} M R^2$) to
reach this limiting situation at $\nu_s\approx 300$ Hz. This
quadrupole is only present during accretion and makes it difficult to
form radio pulsars with $\nu_s>(600-800) \ {\rm Hz}$ by accreting at
$\dot M \gtrsim 10^{-10} M_\odot \ {\rm yr^{-1}}$.  The gravity wave
strength is $h_c\sim (0.5-1) \times 10^{-26}$ from many of these
neutron stars and $>2\times 10^{-26}$ for Sco X-1. Prior knowledge of
the position, spin frequency and orbital periods will allow for deep
searches for these periodic signals with gravitational wave
interferometers (LIGO, VIRGO and the ``dual-recycled'' GEO 600
detector) and experimenters need to take such sources into
account. Sco X-1 will most likely be detected first.

\end{abstract}


\bigskip

\centerline{\bf  To Appear in The Astrophysical Journal Letters}

\bigskip

\section{Introduction} 

The launch of the {\it Rossi X-Ray Timing Explorer} (RXTE) has allowed
for the discovery of fast quasi-periodic variability from many rapidly
accreting ($ \dot M\gtrsim 10^{-11} M_\odot \ {\rm yr^{-1}}$) neutron
stars. These observations strongly suggest that these neutron stars
(NSs) are rapidly rotating, as predicted by those scenarios connecting
the millisecond radio pulsars to this accreting population (see
Bhattacharya 1995 for an overview). Strohmayer et al. (1996) were the
first to detect nearly coherent $\nu_B=363$ Hz oscillations during
type I X-ray bursts from the low accretion rate ($\dot M< 10^{-9}
M_\odot \ {\rm yr^{-1}}$) NS 4U~1728-34. Pulsations were detected in
six of the eight bursts analyzed at that time. In addition, two high
frequency quasi-periodic oscillations (QPOs) were seen in the
persistent emission. These changed with accretion rate, but maintained
a fixed difference frequency of $\nu_d\approx 363 $ Hz, identical to
the period seen during the bursts.  The detection of two drifting
QPO's (in the persistent emission) separated by a fixed frequency
identical to that seen in the bursts naturally leads to beat frequency
models (Strohmayer et al. 1996; Miller, Lamb, \& Psaltis 1998). The
difference frequency is presumed to be the NS spin frequency, $\nu_s$,
whereas the upper frequency has different origins in different models
(see van der Klis 1998 for a summary). In addition, the temporal
behavior of the periodic oscillations both during the rise of the
bursts (Strohmayer, Zhang, \& Swank 1997b) and in the cooling tails
(Strohmayer et al. 1997a) are most easily explained in terms of
rotation.

There are six NSs with measured periodicities during Type I X-ray
bursts (see Table 1). Both the difference frequencies ($\nu_d$) and
the burst frequencies ($\nu_B$) are in a narrow range, from 260
to 589 Hz. For two objects (KS 1731-260 and 4U 1636-53) the difference
frequencies are one-half the burst values. Which value is $\nu_s$ is
not resolved.  There are also many NSs that accrete at higher
rates and are not regular Type I X-ray bursters. Many of these
objects, notably the ``Z'' sources, also show drifting QPO's at fixed
separation, again with a similarly narrow frequency range (roughly
250-350 Hz). Beat-Frequency like models are also applied to these
observations so as to infer $\nu_s$. The applicability of such a model
is less clear when the difference frequency is not constant (Sco X-1,
van der Klis et al. 1997; 4U 1608-52, Mendez et al. 1998). 

If accreting matter always arrives with the specific angular momentum
of a particle orbiting at the NS radius ($R=10 R_6 {\rm km}$), then it
only takes $\sim 10^7 \ {\rm yrs}$ of accretion at $\dot M\approx
10^{-9} M_\odot \ {\rm yr^{-1}}$ for a $M=1.4M_{1.4} M_\odot$ NS to
reach $\nu_s=50$ Hz from an initially low frequency. It is thus
remarkable that these NSs are all rotating at nearly the same
rate. White and Zhang (1997) argued that this similarity arises
because these NSs are magnetic and have reached an equilibrium where
the magnetospheric radius equals the co-rotation radius. This requires
an intrinsic relation between their magnetic dipoles, $\mu_b$, and
$\dot M$ so that they all reach the same rotational equilibrium (most
likely $\mu_b\propto \dot M^{1/2}$) and a way of hiding the persistent
pulse typically seen from a magnetic accretor.

My alternative explanation for these spin similarities is that
gravitational wave (GW) emission has started to play an important
dnrole. If the NS has a misaligned quadrupole moment, $Q$, then the
strong spin frequency dependence of GW emission defines a critical
frequency beyond which accretion can no longer spin-up the star.  Such
a NS will radiate energy via GW's at the rate $\dot E=32
GQ^2\omega^6/5 c^5$, where $\omega=2\pi \nu_s$, and lose angular
momentum at the rate $N_{gw}=\dot E/\omega$. Balancing this spin-down
torque with the characteristic spin-up torque from time-averaged
accretion, $N_a\approx \dot M(GMR)^{1/2}$, gives the $Q$ needed so as
to make the critical frequency 300 Hz,
\begin{equation}\label{eq:qneed} 
Q\approx 4.5 \times 10^{37} \ {\rm g \ cm^2}\left(\dot M\over 10^{-9}
\ {\rm M_\odot \ yr^{-1}}\right)^{1/2}\left(300 \ {\rm Hz}\over
\nu_s\right)^{5/2},
\end{equation} 
or $<10^{-7}$ of the NS moment of inertia, $I\approx 10^{45} \ {\rm g
\ cm^2}$. The similarities in $\nu_s$ may then arise because of the
weak dependencies of the critical frequency on $Q$ and $\dot M$.

What is the source of the misaligned quadrupole?  Wagoner (1984)
argued that accreting NSs would get hung-up at spin frequencies where
the Chandrasekhar-Friedman-Schutz (CFS) instability sets in. However,
Lindblom (1995) and Lindblom \& Mendell (1995) have shown that the
star needs to be very near the breakup frequency ($\nu_s \gtrsim \
{\rm kHz}$) for such an instability to occur, even for the core
temperatures $T_c=(1-3)\times 10^8 \ {\rm K}$ of rapidly accreting NSs
(Ayasli \& Joss 1978; Brown \& Bildsten 1998). The spin frequencies
for these NSs are too slow for such an instability.

I present in \S 2 a new source for lateral density variations in an
accreting NS; electron captures (hereafter EC) in the crust. The
constant compression of the crust forces nuclei to undergo EC when the
electron Fermi energy, $E_F$, is high enough to make a
transition. However, the crust is hot enough in a rapidly accreting
NSs to make the EC rates temperature sensitive. Hotter regions then
capture at lower pressures, so that the density jump associated with
the EC is at a higher altitude in the hotter parts of the crust.
Moderate lateral temperature variations then lead to density
variations large enough to generate the required $Q$. This outcome is
independent of the particular source of the temperature
variations. One possible cause for a $T$ asymmetry relative to the
spin axis is a weak magnetic field. I conclude in \S 3 by finding the
GW signal strength and estimating detection.

\section{Electron Captures in the Neutron Star Crust}

 These NSs accrete hydrogen and helium rich material from their
companions, and within days of reaching the surface, this matter is
burned to heavy elements. The composition of these ashes is still
uncertain, but most certainly consists of heavy nuclei, potentially
beyond the iron group (Schatz et al. 1997).  This material replaces
the primordial NS crust and becomes neutron rich by successive
EC. Later on, neutron emissions and pycnonuclear reactions occur 
(Bisnovatyi-Kogan \& Chechetkin 1979; Sato 1979; Haensel \& Zdunik 1990
(hereafter HZ)). For the $\dot M$'s and residual hydrogen abundance
(Taam, Woosley, \& Lamb 1996) appropriate for these NSs, the
temperature of the accumulating matter and deep crust is $T_8=T/10^8 \
{\rm K} \approx (2-6)$ (Brown \& Bildsten 1998).

 I confine my discussion in this {\it Letter} to the outer crust
(before neutron drip at $\rho\approx (4-6)\times 10^{11} \ {\rm
g \ cm^{-3}}$, HZ), which is held up by degenerate and relativistic
electrons. They exert a pressure $p=1.42\times 10^{30}\ {\rm erg \
cm^{-3}} (E_F/30 {\rm MeV})^4$ from which the mass of the shell above
a particular depth is $M_{cr}(E_F)=4\pi R^2 p/g$, where $g=GM/R^2$, or
\begin{equation}
M_{cr}(E_F)\approx 5\times 10^{-5} M_\odot 
{R_6^4 \over M_{1.4}}\left(E_F\over
30 \ {\rm MeV}\right)^4. 
\end{equation}
The crust is compressed on a timescale $t_{\rm comp}\equiv p/\dot m
g$, where $\dot m$ is the local accretion rate, so that $E_F$ rises
with time, eventually leading to nuclear EC. Many (Sato 1979; HZ)
presumed that the captures were instantaneous once $E_F>E_d$, where
$E_d$ is the mass difference between the (A,Z) nucleus and (A,Z-1)
nucleus. In reality, the transitions will not occur until the EC
lifetime is comparable to $t_{\rm comp}$ (Blaes et al. 1990). For low
accretion rates $\dot M\sim 10^{-16} M_\odot \ {\rm y^{-1}}$, Blaes et
al. (1990) showed that most captures occur in a thin zone where
$E_F>E_d$ only slightly. The situation is 
different for the high $\dot M$'s of the bright X-ray
sources, as the high crustal temperature leads to most EC
occurring out on the thermal tail at a physical location where
$E_F<E_d$.

\begin{figure}[hbp]
\centering{\epsfig{file=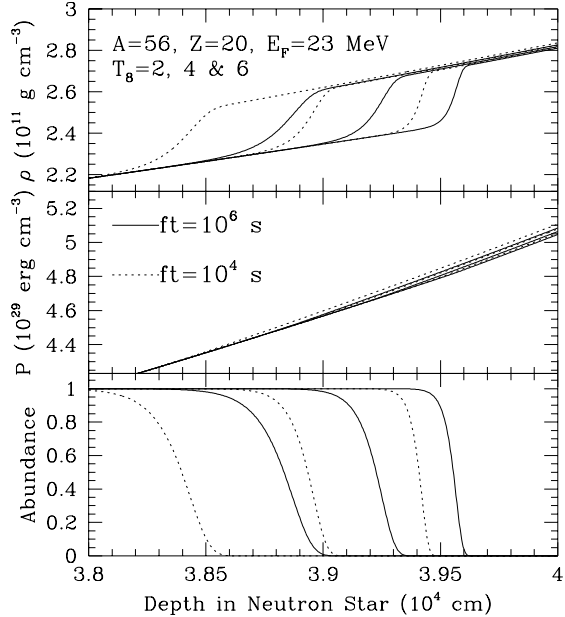,width=3 in}}
\caption{
%
%
%
The density, pressure and nuclear abundance in the $^{56}$Ca electron
capture layer for a $R=10 \ {\rm km}$, $M=1.4 M_\odot$ NS accreting
at $\dot M=2 \times 10^{-9} M_\odot \ {\rm y^{-1}}$.  These
are plotted as a function of increasing depth into the star, deeper
regions are to the right. For a fixed value of $ft$, the hotter crusts
deplete sooner. The curves are, from left to right, for $T_8=6,4 $ and
2. 
\label{fig:capture}}
\end{figure}

 I consider a region of the crust consisting of a single nucleus of
charge $Ze$ and mass $Am_p$. For many nuclei, the capture to the
ground state of the (A,Z-1) nucleus is highly forbidden and proceeds
more slowly than $t_{\rm comp}$. The element is still abundant when
$E_F$ reaches a value where a more favorable transition to an excited
state of the (A,Z-1) nucleus can occur.\footnote{Since the mass
difference of the (A,Z-1) and (A,Z-2) nuclei is smaller than the mass
difference of the (A,Z) and (A,Z-1) nuclei, the sequence always
consists of two successive captures (HZ). This also blocks the inverse
reactions. The larger phase space for the second reaction (as well as
the larger number of available excited states with favorable
spin-parity) means that the (A,Z-1) nucleus captures
electrons more rapidly than the (A,Z) nucleus. Since the first capture
is the rate-limiting step, I do not track the intermediate (A,Z-1)
nucleus.} For this reason, I estimate the EC rates by using $ft$
values in the range of allowed reactions, $ft=10^4-10^6\ {\rm
s}$. Since these stars are hot enough for the sub-threshold EC to
dominate, the transition occurs when $E_{F,depl}=E_d-\beta k_BT$,
where $\beta\approx (8-12)$ depends logarithmically on $\dot m$, $T$
and $ft$ (Bildsten \& Cumming 1998). A hotter region undergoes the
transition at a higher altitude (call $s_c$ the distance down to the
transition from where $E_F=1 \ {\rm MeV}$, i.e. $s$ increases into the
star) compared to a colder region. Figure \ref{fig:capture} displays
the EC transition layer for a NS accreting at $\dot M=2\times 10^{-9}
\ M_\odot \ {\rm y^{-1}}$. As an example I chose the nucleus
$^{56}$Ca, which captures at $E_d\approx 23$ MeV, where $\rho\approx
2.5\times 10^{11} \ {\rm g \ cm^{-3}}$. This is the fourth EC
transition that HZ found when starting with $^{56}$Fe. Since this is a
generic process, I show cases for $ft=10^4 \ {\rm s}$ (dotted) and
$ft=10^6 \ {\rm s}$ (solid) for (from left to right in the upper and
lower panels) $T_8=6,4 $ and 2. The density jump associated with the
ECs is evident and, if discontinuous, would be $\Delta
\rho/\rho=2/(Z-2)=0.11$.

If a transverse temperature gradient is present in the crust, then
these ECs generate density asymmetries on the star and a quadrupole
with mass $\Delta M\approx 4\pi R^2 \Delta \rho \Delta s_c$. The mass
inferred from the $ft=10^4 \ {\rm s}$ case for a temperature contrast
$\Delta T_8=2$ (where the difference in physical depth of the EC is
$\Delta s_c\approx 500 \ {\rm cm}$) is $\Delta M\approx 10^{-7}
M_\odot$, just what is needed for GW's to be important.  More
generally, the mass involved is $\Delta M=M_{cr} (\Delta
\rho/\rho)(4\Delta E_{F, depl}/E_F)$, or since $\Delta E_{F,
depl}\approx 10 k_B \Delta T$,
\begin{equation}\label{eq:mquad}
\Delta M\approx 5.5 \times 10^{-8}  M_\odot \Delta T_8
{R_6^4\over M_{1.4}}\left(E_F\over
30 \ {\rm MeV}\right)^3, 
\end{equation}
for a single EC layer. Since there are a few more electron capture
layers underneath this one, there is adequate mass to generate the
required $Q$ when $\Delta T_8\approx 1$. The repercussions of
the deeper reactions is still unknown.

The EC's deposit heat directly into the crust, generating a flux
$F_{\rm deep}= E_{\rm deep}\dot m$, where $E_{\rm deep}\approx 10^{17}
\ {\rm erg \ g^{-1}}$ (HZ). Temporal variations in $\dot m$ lead to
local heating at these depths, which takes a few years to equilibrate
in the radial direction. The heat transfer is fixed by electron
conduction at these depths, so that the conductivity is $K=\pi^2 k_B^2
T n_e c^2 /3 E_F \nu_{e-ph}$, where $\nu_{e-ph}\approx 13 \alpha k_B
T/\hbar$ is the electron-phonon scattering frequency (Yakovlev \&
Urpin 1980). The time it takes heat to cross a scale height, $H=p/\rho
g$, is then $t_{\rm th}\approx {\rho C_p H^2 /K}\sim {\rm yr}$ where
$C_p\approx 3 k_B/Am_p$ is the specific heat at constant pressure.

 However, lateral thermal asymmetries will persist as long as the
heating time, $t_H\approx C_p T y/E_{\rm deep} \dot m\sim 4\times
10^{-3} T_8 t_{\rm comp}$, is shorter than the time to transport heat
around the star at the same depth.  That time is $t_{\rm th,R}\approx
{\rho C_p R^2/K} \approx 6400 \ {\rm y}(E_F/30 \ {\rm MeV})$, so for
$\dot M\gtrsim  3\times 10^{-11} M_\odot {\rm y^{-1}} (E_F/30 \ {\rm
MeV})^3 (R/10 \ {\rm km})^2$, the time to transport heat around the
crust is longer than the time to locally heat it.  All of the objects
I am discussing are in this regime and so large-scale temperature
asymmetries can persist as long as the system is being perturbed in
some way. Though the origin of lateral $T$ gradients is still unknown,
it is difficult for conduction to wash them out.  For example,
large-scale $\dot m$ or compositional variations (due to, say,
magnetically channeled flow or a ``buried field'') will imprint large
scale $T$ asymmetries.

The pressure as a function of depth in the EC region is shown in the
middle panel of Figure 1.  Even though the density contrasts are
large, the pressure contrasts are not. If one part of the star has a
slightly deeper capture zone, at $s_{c,2}=s_{c,1}+\Delta s_c$, then
the pressures at a fixed depth $s> s_{c,1}$ relate via $P_1\approx
P_2(1+8\Delta s_c/Zs)$. The pressure is higher underneath the layer
that captured sooner due to the ``extra weight'' of the dense
layer. Since $s\propto E_F$, the lateral pressure contrast at a fixed
depth, $\Delta P=(P_1-P_2)$, is $\Delta P/P\approx 80 k_B \Delta
T/ZE_F$.  These transverse pressure gradients would lead to flow and a
cancellation of the quadrupole if the matter was in a liquid
state. However, as long as $T_8< 10$, the matter is solid at these
depths and has a finite shear modulus, $\mu\approx 10^{-2} p$
(Strohmayer et al. 1991). The transverse pressure gradients are then
balanced by a slight amount of shear stress as the matter underneath
the cooler regions laterally shifts a distance $\xi$. Roughly, 
$\xi$ is found from the transverse momentum equation $\Delta
P/R\approx \mu \xi /s^2$, or $\xi/s\approx (100s/R)(8\Delta s_c/Zs)$.
This gives $\xi/s \approx 6\times 10^{-3} \Delta T_8$, or a few meters
of transverse displacement over the whole surface. For now, I will
presume that this configuration will not crack. Relative
vertical motion has yet to be investigated and could reduce the
estimated $Q$.

\section{Detectability of the Gravitational Waves}

Most accreting, weakly magnetic neutron stars are spinning at
$\nu_s\approx (250-500) \ {\rm Hz}$. I conjecture that this similarity
is the result of an equilibrium where the angular momentum accreted is
radiated away in gravitational waves. I showed that pre-threshold
electron captures will turn any lateral temperature gradients into
lateral density gradients. In this case, a small spin-misaligned
temperature gradient gives rise to a quadrupole, $Q\sim
10^{-7} MR^2$, adequate to explain the similarities at $\approx 300 \
{\rm Hz}$.  These accretion-induced $Q$'s clearly make it
difficult to spin up a neutron star to 1000 Hz by accreting at $\dot M
\gtrsim 10^{-10} M_\odot \ {\rm yr^{-1}}$. Indeed, a NS spinning at
1000 Hz with $Q\approx 10^{-8} MR^2$ would be
difficult to spin-up even at $\dot M\sim 10^{-8} M_\odot \ {\rm
yr^{-1}}$. The lateral $T$
asymmetries should subside at lower $\dot M$'s, reducing  $Q$ 
(see the dependence in equation [\ref{eq:mquad}]) and
potentially making it easier to reach higher $\nu_s$ by accreting
slowly for a long period of time. There is already some hint of this,
as the highest $\dot M$ sources seem to be rotating the slowest.

Independent of the cause of the $T$ asymmetry, if gravitational
radiation is the explanation of the spin similarities, then the
characteristic GW signal strength at Earth (suitably averaged over
spin orientations, Brady et al. 1998) is $h_c=2.9 G\omega^2 Q/dc^4$,
where $d$ is the distance to the object (the prefactor is 4 when
looking down the spin axis). Presuming the NS luminosity is $L\approx
GM\dot M/R$ then $h_c$ is written in terms of the observable $F=L/4\pi
d^2$ (Wagoner 1984)
\begin{equation} \label{eq:hsecond}
h_c\approx 4 \times 10^{-27}{R_6^{3/4}\over M_{1.4}^{1/4}}
\left(F\over 10^{-8} \ {\rm erg \ cm^{-2} \ s^{-1}}\right)^{1/2}
\left(300 \ {\rm Hz}\over \nu_s\right)^{1/2},  
\end{equation}
which represents a lower limit for $h_c$ due to the $L$ to $\dot M$
conversion I chose and since $F$ is never fully measured. The
minimum $h_c$'s for those NSs which have $\nu_s$ inferred
from Type I bursts are shown in Table 1. There are
many NSs for which only $\nu_d$ has been measured (see van der Klis
1998 for a summary). The average 2-10 keV fluxes from van Paradijs
(1995) imply the following GW amplitudes: GX 349+2 ($\nu_d=266\pm 13$,
$h_c\approx 5.4\times 10^{-27}$, Zhang et al. 1998b), 4U~1820-30
($\nu_d=275\pm 8$, $h_c\approx 3.7\times 10^{-27}$), GX 17+2
($\nu_d=294\pm 8$, $h_c\approx 4.7\times 10^{-27}$), 4U~0614+09
($\nu_d=327\pm 4$, $h_c\approx 1.3\times 10^{-27}$), GX 5-1
($\nu_d=327\pm 11$, $h_c\approx 6.0\times 10^{-27}$) Cyg X-2
($\nu_d=343\pm 21$, $h_c\approx 3.7\times 10^{-27}$), and GX 340+0
($\nu_d=325\pm 10$, $h_c\approx 3.7\times 10^{-27}$, Jonker et
al. 1998). 

The highest flux object that shows kHz QPO's is Sco X-1 with an
orbital period of 18.9 hours. It's difference frequency is not
constant however (van der Klis et al. 1997) and so we are still
uncertain about the exact spin frequency. If it is rotating at
$\approx 250 \ {\rm Hz}$, then $h_c\approx 2.2\times 10^{-26}$ at
$2\nu_s=500 \ {\rm Hz}$. Detection of a weak periodic gravitational
wave signal depends on the ability to coherently fold large data sets
for a long time, $\tau$, in order to obtain the $\sim (2 \nu_s
\tau)^{1/2}$ signal enhancement. The initial LIGO interferometer will
have $h_{rms}=10^{-21}$ at $\nu\approx 500 \ {\rm Hz}$ (Abramovici et
al. 1992), requiring integration times $\sim \ {\rm yr}$.  If the
orbital parameters, spin frequency and phase were known for Sco X-1,
then Brady et al.'s (1998) work implies it would be detected at 99\%
confidence with the initial LIGO in about one year. However, our lack
of detailed prior information on these parameters demands a larger
parameter space search. The consequent reduction in the $S/N$ (maybe a
factor of $\sim 5$) make it more likely that detection will have to
wait for the enhanced LIGO (see Thorne 1998 for the
sensitivities). The similar spin frequencies of so many objects give
an advantage to dual-recycling interferometers (such as GEO~600 and
TAMA 300). Indeed, the GEO 600 noise strength (from Strain et
al. 1998) when tuned to Sco X-1 is competitive with the initial LIGO
and VIRGO. Detection of the larger number of NSs\footnote{The recent
RXTE discovery of a persistent 401 Hz accreting pulsar (Wijnands \&
van der Klis 1998) in the transient binary SAX J1808.4-3658 (in't Zand
et al. 1998) provides hope for GW studies. It was observed in outburst
by RXTE (Marshall 1998) and it's orbit measured by Chakrabarty \&
Morgan (1998).  Enough is known about this weakly magnetic ($B < 10^9
\ {\rm G}$) accreting pulsar that a coherent data fold can be
undertaken. Unfortunately, the time averaged flux (I estimate
$2\times 10^{-10} \ {\rm ergs \ cm^{-2} \ s^{-1}}$ at most) is low and
even if looking right down the spin-axis, $h< 7\times 10^{-28}$. It
might be that spin-down in quiescence is the best indicator of GW
emission from this NS.} with $h=(0.5-1)\times
10^{-26}$ will have to wait for
advanced broad-band interferometers (such as advanced LIGO), further
developments in the dual-recycled interferometers, or larger cryogenic
detectors.

The stability of the GW signal depends on the origin of the mass
quadrupole. For the EC origin, I would not expect major differences in
$Q$ on timescales shorter than the thermal time (years).  Another
relevant time is that to replenish the thickness of the EC layer,
which is $\sim \ 100$ years at $\dot M\sim 10^{-9} M_\odot \ {\rm
yr^{-1}}$. However, since $\dot M$ varies, we cannot expect the
torques to always be in balance. For example, if $\dot M$ were to drop
suddenly, the instantaneous spin-down of the NS due to the equilibrium
$Q$ would be $\dot \nu \approx 10^{-13} \ {\rm s^{-2}} (\langle \dot
M\rangle/10^{-9} M_\odot \ {\rm yr^{-1}})$. This would lead to one
pulse cycle of drift in a time $\approx (2/\dot \nu)^{1/2}\approx 50 $
days.

How does this work affect the accretion spin-up scenario for making
millisecond pulsars? The EC $Q$ will decrease once accretion has
halted and the temperature equilibrates. In addition, when the crust
is cold, it only takes $10^{-7} M_\odot$ of accretion to wash out the
compositional gradients generated during accretion at high
rates. There is an important comparison to make in this
regard. Namely, as long as $Q$ decays on a timescale, $t_d$, much
shorter than $t_{\rm spin-up}\equiv I\omega_i/ \dot M
(GMR)^{1/2}\approx 5\times 10^7 {\rm yr}$ then the millisecond pulsar
is ``born'' at roughly the same spin period since $\nu_f=\nu_i(1 + 4
t_d/t_{\rm spin-up})^{-1/4}$. If $Q$ persists at some low level long
after accretion has ended, it could still play an important
role. Indeed, as many have shown (Thorne 1987, New et al. 1995, Brady
et al. 1998) a small $Q$ can go a long way towards explaining the
observed spin-down of millisecond pulsars {\it and} making them
detectable as GW sources.

Transverse $T$ gradients may be more prevalent on the $B\gtrsim
10^{11} \ {\rm G}$ accreting pulsars. However, most of them have
$\nu_s \ll \ {\rm Hz}$ so that the $Q$ needed for GW equilibrium is
implausibly large, making it unlikely that GW's play an important role
in their spin evolution.  However, there is a lack of accreting
pulsars with $\nu_s > \ {\rm Hz}$ that is usually explained by stating
that all pulsars have reached their magnetic equilibrium and that few
have $B\ll 10^{11} \ {\rm G}$.  If we presume a maximum allowed
quadrupole $Q_{max}=10^{-5} MR^2$ (Thorne 1987), then GW emission
could play an important role when $\nu_s > 35 \ {\rm Hz}(\dot M/
10^{-9} \ M_\odot \ {\rm yr^{-1}})^{1/5}$, only somewhat alleviating
this issue.

I thank E. Brown, A. Cumming, E. Flanagan, T. Prince, H. Schatz,
G. Ushomirsky and I. Wasserman for helpful discussions and J. Arons,
R. Rutledge, and M. van der Klis for comments on the
manuscript. P. Brady and K. Thorne clarified issues of GW detection.
This research was supported by NASA via grants NAG 5-2819 and
NAGW-4517, a Hellman Family Faculty Fund Award (UCB) and the Alfred
P. Sloan Foundation.

\begin{table}[htb]
\begin{center}
\caption{Rapid Periodicities During Type I X-Ray Bursts}
\begin{tabular}{llllll}
\hline 
Object Name & $\nu_B$ & $\nu_{d}$ & Flux$^a$ & $h_c$ & Ref.$^b$\\
  & (Hz) & (Hz) &  & ($10^{-27}$) & \\
\hline
4U 1702-429 & 330 & -- & 1.0 & 1.2 & [1]\\
4U 1728-34 & 363 & 363 &2.8 & 2.0 &  [2] \\
KS 1731-260 & 524& $260\pm 10$& 0.2-2 &2.0  & [3-4] \\
Aql X-1 & 549 & -- & &  & [5]\\
4U 1636-53 & 581 & $276\pm 10$  & 4.4 & 2.8  & [6-7] \\
MXB 1743-29 & 589 & -- & && [8]\\
\hline
\end{tabular}
\end{center}
\noindent
$^a$ Average 2-10 keV fluxes (in units $10^{-9} \ {\rm erg \ cm^{-2} \
s^{-1}}$) are from  van Paradijs (1995). 
 
\noindent 
$^b$ REFERENCES: [1] Swank et al. 1997, [2] Strohmayer et al. 1996, [3] Smith et
al. 1997, [4] Wijnands \& van der Klis 1997, [5] Zhang et al. 1998a,
[6] Strohmayer et al. 1998, [7] Wijnands et al. 1997, [8] Strohmayer et
al. 1997a
\end{table}
 
\end{document}